\documentclass[hyper,11pt,letterpaper]{JHEP3}
\usepackage[dvips]{epsfig}
\usepackage{amsmath,amssymb,epsf,amsfonts}
\usepackage{graphicx}
\usepackage{dcolumn}
\usepackage{bm}

\def\<{\langle}

\def\>{\rangle}
\def\({\left (}
\def\){\right )}
\def\[{\left[}
\def\]{\right]}

\def\beq{\begin{equation}}
\def\eeq{\end{equation}}

\newcommand{\bea}{\begin{eqnarray}}
\newcommand{\eea}{\end{eqnarray}}

\def\tone{\theta_1}
\def\ttwo{\theta_2}
\def\tthree{\theta_3}
\def\tfour{\theta_4}

\author{Andreas Karch,\!$^1$\footnotemark[1]\,
Piotr Sur\'owka,\!$^{1,2}$\footnotemark[2]\, and Ethan G.
Thompson\!$^1$\footnotemark[3]
\\
$^1$Department of Physics, University of Washington, Seattle, WA
98195-1560
\\
$^2$Institute of Physics, Jagiellonian University, Reymonta 4,
30-059 Krak\'ow, Poland }

\footnotetext[1]{E-mail: \email{karch@phys.washington.edu}}
\footnotetext[2]{E-mail: \email{surowka@u.washington.edu}}
\footnotetext[3]{E-mail: \email{egthomps@u.washington.edu}}

\abstract{We study defects in non-relativistic conformal field
theories. As in the well-studied case of relativistic conformal
defects, we find that a useful tool to organize correlation functions
is the defect operator expansion (dOPE). We analyze how the dOPE is
implemented in theories with a holographic dual, highlighting some
interesting aspects of the operator/state mapping in
non-relativistic holography. }
\title{A holographic perspective on non-relativistic
conformal defects}
\preprint{INT-PUB-09-013}

\begin{document}

\section{Introduction}\label{intro}

Experimentalists have made great progress in creating fermionic
superfluids in the laboratory.
 An important tool in these experiments is the Feshbach resonance
applied to ultracold atomic gases.  Using these systems, the BCS
superfluid, the unitary Fermi gas, and the Bose-Einstein condensate
have all been
created\cite{Ketterle:2008, Giorgini:2008zz, Bloch:2008zzb}. However,
in all of these situations the fermion pairs form in the
\textit{s}-wave channel rather than the \textit{p}-wave channel.
Experiments have shown that \textit{p}-wave Feshbach bound states
are unstable.  It would therefore be of great interest to find a
different mechanism for the production of \textit{p}-wave fermionic
superfluids.

Such an alternative mechanism was recently proposed in
\cite{Nishida:2008gk}. This paper presents a model of two
interacting fermionic species A and B, where the A-type fermions are
confined to live on a two-dimensional defect, while the B-type
fermions fill the entire three-dimensional space.  The two species
interact via the conventional \textit{s}-wave Feshbach resonance,
which induces an attractive interaction among the defect fermions.
Because the defect fermions are identical and therefore have
identical spin state, this attractive force leads to \textit{p}-wave
pairing. This system may be experimentally realizable using lithium,
potassium and optical traps.  In \cite{Nishida:2008gk}, several
properties of this system are calculated in the weak-coupling limit.

In the unitarity limit, a system of trapped fermionic atoms (such as the one
featured in \cite{Nishida:2008gk}) becomes
scale-invariant and enjoys the full Schr\"odinger symmetry algebra.
Such a field theory is referred to as a non-relativistic conformal
field theory (NRCFT).  Fermions at unitarity are an inherently
strongly-coupled system and it is thus difficult to calculate their properties.
A solvable toy model of strongly coupled NRCFTs has
recently been developed in the form of a holographic description of
field theories with the Schr\"odinger symmetry \cite{Son:2008ye, Balasubramanian:2008dm}.  In analogy to
the familiar AdS/CFT correspondence, these holographic models are
tractable on the bulk side when the field theory is strongly-coupled.
These models should be useful to study aspects of NRCFTs
that are essentially determined by the symmetries
and provide a useful laboratory to build intuition
about the behavior of these strongly-coupled systems.

Similarly, one would hope that a holographic description of non-relativistic defect CFTs (NRdCFTs) may offer some insight into the
models introduced by \cite{Nishida:2008gk} for
\textit{p}-wave fermionic superfluids. In this paper, we present
some initial investigations into holographic descriptions of
NRdCFTs. In particular, we will argue that a very useful tool in the
analysis of NRdCFTs is the defect operator expansion (dOPE),
which is already familiar from the relativistic setting. A
holographic description of the dOPE has been carried out for relativistic defect
CFTs in \cite{Aharony:2003qf}.  This material is reviewed in Section
\ref{cftd}.  We find that many features of relativistic defect CFTs (dCFTs)
carry over to the non-relativistic case, with one crucial new
feature in the non-relativistic scenario.  As shown in Section
\ref{nrdope}, the defect Schr\"odinger geometry allows for the
presence of an arbitrary function of the radial coordinate.  This
function characterizes different NRdCFTs, and is related to
the `extra' lightcone dimension present in Schr\"odinger holography
as compared to relativistic holography.

In Section \ref{holo}, a specific example of an NRdCFT is
constructed by applying the Null Melvin Twist (NMT) to the Janus
solution \cite{Bak:2003jk} of type IIB supergravity.
This solution represents a particular choice of the arbitrary
function mentioned above, and gives rise to a defect theory without
matter localized to the defect.

We present possible future directions of this research and our conclusions in Section \ref{conclude} and give more
details of our calculations in the appendices.

\section{Conformal field theories with defects}
\subsection{Review of correlators in relativistic dCFTs}\label{cftd}

Introducing a codimension one defect into a relativistic conformal field theory
obviously breaks some of the spacetime symmetries of the theory, such as translation and
rotation invariance in the directions transverse to the defect.
Conformal defects preserve an $SO(d-1,2)$ subgroup of the $SO(d,2)$
conformal symmetry of the $d$-dimensional CFT that includes scale
transformations. Of course, this is just the conformal group in $d-1$
dimensions. Translations, boosts and rotations within the defect act
in the natural way on the $d-1$ coordinates along the defect (which
we will refer to as
 $\vec{x}$ and
$t$). What distinguishes a $d$-dimensional dCFT from a ($d-1$)-dimensional CFT is that the scale transformations (and also the
inversion) also act on the coordinate transverse to the defect
(which we will refer to as $y$). In particular, the dilatation
rescales $\vec{x} \rightarrow \lambda \vec{x}$, $t \rightarrow
\lambda t$ as well as $y \rightarrow \lambda y$.

Since the symmetry is reduced, correlation functions in a dCFT are
much less constrained than correlation functions in a CFT. The constraints have
been worked out in full generality in \cite{McAvity:1995zd}. For
example, in a dCFT, scalar operators can have non-trivial one point
functions. In a CFT without defect, translation invariance requires the one-point
function of any operator other than the identity to vanish. In a dCFT, the dependence on $y$ is not restricted
by translation invariance, so scalar operators can have a one-point
function $\langle O(t,\vec{x},y) \rangle= A_O y^{- \Delta}$. Two-point functions of scalar operators $O_1$ and $O_2$ of dimension
$\Delta_1$ and $\Delta_2$ respectively can even depend on an
arbitrary function $f$ \beq \label{twopt} \langle  O_1(t,\vec{x},y)
O_2(t',\vec{x}',y') \rangle = y^{-\Delta_1} (y')^{- \Delta_2}
f(\xi), \eeq where $\xi$ is the conformally invariant variable \beq
\xi = \frac{(x^{\mu} -(x')^{\mu}) (x_{\mu} - (x')_{\mu})}{4 y y'}.
\eeq

A powerful tool in studying dCFTs is the defect operator expansion (dOPE). It allows one to expand any operator
$O(t,\vec{x},y)$ in the full $d$ dimensional dCFT (which, following
\cite{Aharony:2003qf} we'll refer to as an ambient space operator)
in terms of defect localized operators $\hat{O}_n(t,\vec{x})$ with
scaling dimension $\hat{\Delta}_n$. The $SO(d-1,2)$ defect conformal
group acts on these defect localized operators as the standard
conformal group of a $d-1$ dimensional non-defect CFT, so the
correlation functions of the $\hat{O}_n$ operators obey the standard
non-defect CFT constraints \cite{Osborn:1993cr}. The constraints on
the correlation functions of the full dCFT, like the one-point and two-point
functions of scalar operators quoted above, can then be understood as
a consequence of the dOPE together with the standard ($d-1$)-dimensional non-defect CFT constraints on the correlation functions
of the $\hat{O}_n$.

The dOPE of a scalar operator reads \cite{McAvity:1995zd} \beq
\label{dope} O(t,\vec{x},y) = \sum_n \frac{B_O^{\hat{O}_n}}{
y^{\Delta - \hat{\Delta}_n}} \hat{O}_n (t,\vec{x}). \eeq In terms of
the dOPE, the one-point function of the operator $O$ in the full dCFT can
be understood as the coefficient $B_O^1$ of the identity operator in
the dOPE, since this is the only operator which can have a
non-vanishing expectation value in the $d-1$ dimensional CFT with
$\langle 1 \rangle =1$. Similarly, the free function $f$ appearing
in the scalar 2-point function can be expanded as \beq f(\xi) =
\sum_{n,m} B_O^{\hat{O}_n} B_O^{\hat{O}_m} y^{\hat{\Delta}_n} (y')^{
\hat{\Delta}_m} \langle \hat{O}_n(t,\vec{x}) \hat{O}_m (t,\vec{x})
\rangle. \eeq The standard constraints on
correlation functions in a $d-1$ dimensional CFT \footnote{ In a
Lorentzian signature CFT one has to be careful about the correct $i
\epsilon$ prescription when using the results of
\cite{Osborn:1993cr}. As explained, for example, in
\cite{Hofman:2008ar}, for a time-ordered two-point function of $O$ with
itself one has \beq \langle T ( \hat{O}(t, \vec{x}) \hat{O}(0,0) )
\rangle = \frac{1}{( -t^2 + |\vec{x}|^2 + i \epsilon)^{\Delta}} \eeq
whereas a non-ordered two-point function would be \beq \langle
\hat{O}(t, \vec{x}) \hat{O}(0,0)
 \rangle = \frac{1}{( -(t - i \epsilon)^2 + |\vec{x}|^2
)^{\Delta}}. \eeq } ensure that the right hand side indeed only
depends on $\xi$.

The simplest example of a dCFT that proved to be useful for
studying the dual gravitational description is the ``no-brane case'' \cite{Aharony:2003qf}. In a CFT
without defect one can promote the line $y=0$ to a defect and interpret the usual
continuity of all fields and their first derivatives as boundary
conditions imposed at the defect. In this case the dOPE just becomes
a standard Taylor expansion, that is $\hat{O}_n = \partial_y^n O$
with $\hat{\Delta}_n = \Delta + n$.

\subsection{The non-relativistic dOPE}
\label{nrdope}

The constraints imposed on correlation functions in a $d-1$
dimensional non-relativistic conformal theory (by which we still
mean time plus $d-2$ spatial dimensions) without defect are quite
different from those in a relativistic theory. For example, the two-point
function between two scalar operators reads
\cite{Henkel:1993sg,Nishida:2007pj} \beq \label{nrtwo} G(t,\vec{x})
= \left \langle T \left ( O(t,\vec{x},y) O^{\dagger}(0,0) \right )
\right \rangle = C t^{-\Delta} \exp \left ( - i N_0
\frac{|\vec{x}|^2}{2t} \right ). \eeq In this expression, $\Delta$
is the dimension of the operator $O$, defined as in the relativistic
case as the eigenvalue under rescalings $D$,
\beq [ D, O(0)] = i \Delta O(0). \eeq

The non-relativistic conformal
algebra has a central element $N$, the particle number.
Correspondingly, scalar operators are not just classified by their
scale dimension, but in addition one needs to specify their particle
number \cite{Nishida:2007pj} \beq [ N, O(0)] = N_O O(0). \eeq The
two-point function depends on both $\Delta$ and $N_O$. In the dual
gravitational description this extra quantum number plays a crucial
role, as we will demonstrate in the next section.

Correspondingly, correlation functions in an NRdCFT differ significantly from their relativistic counterparts.
In \cite{Nishida:2008gk} it was shown that the two-point function of two
defect localized operators $\hat{O}(t,\vec{x})$
takes exactly the same form (\ref{nrtwo}), as expected. However, the two-point function
of two ambient space operators $O(t,\vec{x},y)$ can again depend on
a free function $f(\xi_{NR})$ where\footnote{To see that $\xi_{NR}$
is invariant under the non-relativistic conformal group note that
the most general transformation can be written as
\cite{Henkel:1993sg} \beq t \rightarrow \frac{\alpha t +
\beta}{\gamma t + \delta}, \,\,\,\,\, \vec{x} \rightarrow \frac{
{\cal R} \vec{x} + \vec{v} t + \vec{a} }{\gamma t + \delta},
\,\,\,\,\, y \rightarrow \frac{y}{\gamma t + \delta} \eeq where
${\cal R}$ is a rotation matrix and $\alpha \beta - \gamma
\delta=1$. Under this transformation one gets \beq (t-t')
\rightarrow
 \frac{(\alpha t + \beta)(\gamma t'+\delta) -
(\alpha t' + \beta) (\gamma t + \delta)}{(\gamma t + \delta)(\gamma
t' + \delta)}=
 \frac{t -t'}{(\gamma t + \delta)(\gamma t' + \delta)}.
\eeq } \beq \xi_{NR}=\frac{(t-t')}{yy'}. \eeq Still, the dOPE is
just as constrained in the NRdCFT as it was in the relativistic
dCFT. Time and space translation invariance along the defect
prevent $t$ and $\vec{x}$ from appearing explicitly in the
expansion coefficients, which therefore must be purely functions of
$y$. Scale invariance then fixes the NRdOPE to have exactly the same
form (\ref{dope}) as the relativstic dOPE. As a corollary, the one-point
function in the NRdCFT will also be given by the relativistic
expression  $\langle O(t,\vec{x},y) \rangle= A_O y^{- \Delta}$. The
upshot is that {\it the reduction of correlation functions of
ambient operators to correlation functions of defect localized
operators via the dOPE is identical in the NRdCFT and the
relativistic dCFT}. The correlators of the defect localized
operators are then governed by the corresponding standard
expressions for a non-defect CFT (relativistic and non-relativistic
respectively) in $d-1$ spacetime dimensions.

\section{The dOPE, relativistic and non-relativistic, as a mode expansion}
\subsection{Gravitational description of relativistic dCFTs}
\label{cftgrav}

The AdS/CFT correspondence provides a description of certain
$d$ dimensional CFTs in terms of $d+1$ dimensional gravitational
duals. The $SO(d,2)$ conformal group of the field theory gets mapped to the isometry group of
the dual spacetime metric. Up to the curvature radius, which
encodes the coupling constant of the CFT, this fixes the dual
background uniquely to be AdS$_{d+1}$. Relativistic dCFTs only have
an $SO(d-1,2)$ symmetry, so the dual $d+1$ dimensional metric has
more freedom. The $SO(d-1,2)$ can be made manifest by foliating the
$d+1$ dimensional spacetime with AdS$_d$ slices,
 \beq
\label{warpedads} ds^2 = \frac{e^{2 A(r) }}{z^2} (- dt^2 +
d\vec{x}^2 + dz^2) + dr^2,
\eeq
where as in the last section
$\vec{x}$ denotes the $(d-2)$ spatial coordinates along the defect.
The warpfactor $A(r)$ is completely undetermined by symmetry
considerations. Of course, for any particular dCFT/gravity dual pair,
$A(r)$ is determined by solving the equations of motion in the
bulk\footnote{Following \cite{Aharony:2003qf}, we use the standard AdS/CFT terminology and refer to the
generic points of the $d+1$ dimensional spacetime of the
gravitational dual as the bulk. In contrast, the spacetime on which the field theory
lives is referred to as the defect and its ambient space.}. One
example is the Janus solution of \cite{Bak:2003jk}, which is a
solution of type IIB supergravity.

Relativistic AdS/CFT pairs come with an operator/field
correspondence where every operator on the boundary is dual to a
field in the bulk. Let us for now focus on operators that are dual
to a scalar fields $\phi_{d+1}(t,\vec{x},z,r)$ in the $d+1$
dimensional bulk. Using a separation of variables ansatz,
 \beq
\label{ansatz} \phi_{d+1}(t,\vec{x},z,r) = \sum_n \psi_n(r)
\phi_{d,n}(t,\vec{x},z),
\eeq
one can decompose the bulk scalar into
defect-localized bulk modes $\phi_{d+1}$ (that is, they only depend
on the coordinates of the defect and the radial coordinate of the
bulk), each of which is an eigenfunction of the AdS$_d$ Laplacian
$\nabla_d^2$ acting on the slice coordinates $t$, $\vec{x}$ and $r$:
\beq \label{laplonslice} \nabla_d^2 \phi_{d,n}(t,\vec{x},z) = m_n^2
\phi_{d,n} (t,\vec{x},z). \eeq It was shown in \cite{Aharony:2003qf}
that this mode decomposition in the bulk is the dual of the dOPE of
the dual dCFT operator on the boundary. While we will mostly focus on free
scalar fields, it was argued in \cite{Aharony:2003qf} that such a
decomposition will always be possible, as any bulk field of mass
$M_0$ transforming in some representation of $SO(d,2)$ will
decompose into a tower of $AdS_d$ modes inhabiting representations
of the preserved isometry group $SO(d-1,2)$. As the modes
$\phi_{d,n}(t,\vec{x},z)$ satisfy a standard scalar wave equation
with mass squared $m_n^2$, they are naturally dual to the defect
localized operators $\hat{O}_n (t,\vec{x})$ appearing in the dOPE
(\ref{dope}) with the dimensions given by the standard AdS/CFT
relation,
 \beq
 \hat{\Delta}_n = \frac{d-1}{2} + \frac{1}{2}
\sqrt{(d-1)^2 + 4 m_n^2}.
\eeq
 The $m_n^2$ are the
eigenvalues of the radial equation that is obtained from solving the
equations of motion for $\phi_{d+1}$ that follow from the ansatz
(\ref{ansatz}). For a free scalar field that is \beq \psi_n''(r) + d
A'(r) \psi_n'(r) + e^{- 2 A(r)} m_n^2 \psi_n(r) - M_0^2 \psi_n(r)
=0. \eeq This will receive corrections from various interactions.
An important point to note is that already for a free
scalar field the spectrum of dimensions $\hat{\Delta}_n$ is not
determined from the dimension $\Delta=\frac{d}{2} + \frac{1}{2}
\sqrt{d^2 + 4 M_0^2}$ of the ambient operator $O$ dual to
$\phi_{d+1}$ alone, but the dimensions also depend on the
warpfactor $A(r)$. The coefficient in the dOPE can be extracted from
the wavefunctions $\psi_n(r)$ (see \cite{Aharony:2003qf} for details
and an explicit demonstration that this procedure recovers the
standard Taylor series in the ``no-brane'' case of pure AdS$_{d+1}$
foliated by AdS$_d$ slices).

To verify that the mode decomposition indeed reproduces the dOPE one
can study the behavior of the full field $\phi_d(t,\vec{x},z,r)$ in
the limit of large $r$. The component of the boundary of the
spacetime that is obtained by going to $r \pm \infty$ is an ambient
space point, so asymptotically the spacetime should approach
AdS$_{d+1}$, as locally the ambient space theory is a $d$
dimensional  non-defect CFT. In terms of the warpfactor this means
that $A(r)$ has to asymptotically approach $r$, as the metric
(\ref{warpedads}) with $e^A=\cosh(r)$ is just AdS$_{d+1}$. With this
choice the
asymptotic AdS$_{d+1}$ geometry can be brought into standard flat
slicing form \beq \label{flatslicing} ds^2 = \frac{1}{\tilde{z}^2}
(-dt^2 + d \vec{x}^2 + dy^2 + d \tilde{z}^2) \eeq by the change of
coordinates \beq \label{cov}
 y= z \tanh(r) \rightarrow z,
\,\,\,\,\,\,\,\,\,\,\, \,\,\,\,\,\,\,\,\,\,\, \frac{1}{\tilde{z}} =
\frac{\cosh(r)}{z} \rightarrow \frac{e^r}{2 z} \eeq The standard
AdS/CFT relations give the vacuum expectation value (even in the
presence of sources due to the insertions of other operators) of the
ambient operator $O$ as the coefficient of $\tilde{z}^{\Delta}$ in
$\phi_{d+1}(t,\vec{x},y,\tilde{z})$ in the limit $\tilde{z} \rightarrow 0$. On the
other hand, expanding mode by mode the $\phi_n$ behave at large $r$
and small $z$ as $\langle \hat{O}_n(t,\vec{x}) \rangle e^{-\Delta r}
z^{\hat{\Delta}_n}$. Using the asymptotic change of variables
(\ref{cov}) one can see that the vacuum expectation values of
$\hat{O}_n$ contributes\footnote{One subtlety explained in detail in
\cite{Aharony:2003qf} we are glossing over her is that at large $r$
we don't just get contributions from the leading small $z$ behavior
of the modes $\phi_d$ (which is entirely determined by its
dimension) but in fact all powers of $z$ contribute. Solving the
equations of motion for $\phi_d$ recursively in a power series in
$z$ one can identify these higher powers of z as the contributions
of the descendents of the primaries $\hat{O}_n$ to the dOPE.} to the
expectation values of the ambient operator $O$ exactly with the
coefficient $y^{\hat{\Delta}_n - \Delta}$ as it appears in the dOPE
(\ref{dope})

\subsection{Gravity dual to non-relativistic CFTs}
\label{nrcftgrav}

For a relativistic conformal field theory with gravity dual, the
spacetime geometry was essentially fixed to be AdS by the symmetries
alone, up to the overall curvature radius. Metrics that
exhibit the non-relativistic conformal group (often referred to as
the Schr\"odinger group) as their isometries were analyzed in
\cite{Son:2008ye, Balasubramanian:2008dm}. It was found that the only
way to realize the symmetries of a $d$-dimensional NRCFT was to have a dual
spacetime with {\it two} extra dimensions\footnote{ As in
previous sections we will refer to the $d$ spacetime coordinates as
$t$, $\vec{x}$ and $y$ where we singled out one of the $d-1$ spatial
coordinates as $y$ since it will play the role of a coordinate
transverse to the defect once a defect is introduced.}. The
corresponding metric, which now is usually referred to as the
 Schr\"odinger metric Sch$_{d+2}$, is
 \beq
\label{schmetric}
 ds^2= -\frac{dt^2}{\tilde{z}^4} +
\frac{1}{\tilde{z}^2} ( -2 dv dt  + d \vec{x}^2 + d\tilde{z}^2 +
dy^2)
 \eeq
In this case, there are actually two independent structures in the
metric allowed by the symmetries. The $\frac{dt^2}{\tilde{z}^4}$
piece all by itself is symmetric under the full Schr\"odinger group,
as is the rest of the metric (which is just AdS$_{d+2}$ in
lightcone coordinates). The constant coefficient of
$\frac{dt^2}{\tilde{z}^4}$ can always be scaled to unity by
rescaling $t$ and $v$, but the freedom to have two independent
metric structures that respect the full symmetry will be important
when we generalize to the defect case. Later it was shown that
Sch$_5$ $\times$ $S^5$ can actually be found as a solution to type
IIB supergravity \cite{Herzog:2008wg,Maldacena:2008wh,Adams:2008wt}.

As we pointed out above, one important aspect of this duality is
that the gravitational dual actually has two extra dimensions; that
is, a $d$-dimensional NRCFT (again, we uniformly refer to a theory with $d-1$ space
plus one time dimension as a $d$ dimensional theory) is dual to
Sch$_{d+2}$. The extra lightlike direction $v$ is needed to
compensate the transformations of $d\vec{x}^2 + dy^2$ under Galilean
boosts. Typically $v$ is compact and the inverse radius corresponds to
the mass of the basic particles described by the NRCFT (and, by
convenient choice of units, can be set to unity). The momentum $M$
in the $v$ direction on the gravity side corresponds to the conserved particle number in the field theory.
As a consequence, the operator/field dictionary is significantly
modified in non-relativistic gauge/gravity duality.

As we
mentioned before, operators in an NRCFT are typically taken to have
a fixed particle number, see e.g. \cite{Nishida:2007pj}.
To relate this to the more familiar concept of states
in non-relativistic quantum mechanics having a fixed particle number, one
can note that
NRCFTs have an
operator/state map that relates the dimension of an operator of particle
number $N_{0}=M$ to the
energy of a state in a harmonic trap with the same
particle number \cite{Nishida:2007pj}. On the gravity side any field $\phi_{d+2}(t,\vec{x},y,v,\tilde{z})$
can be decomposed into plane waves along the extra $v$ dimension
\beq
\phi_{d+2}(t,\vec{x},y,v,\tilde{z}) = e^{i \frac{M}{2\pi} v} \phi_{d+1}
(t,\vec{x},y, \tilde{z}).
\eeq
The momentum $M$ maps to particle
number in the field theory. This way the operator/field map
associates to any given field in the bulk not just one operator, but
an infinite tower of operators with particle number $M=0,1,2,3,
\ldots$ Using the wave equation for a free scalar
field\footnote{Note that the only non-trivial components of the
inverse metric are $g^{vv}=1, g^{vt} = g^{tv}=-z^2,
g^{xx}=g^{yy}=g^{zz}=z^2$. This way the Laplacian on Sch$_{d+2}$ is essentially equal
to the Laplacian on AdS$_{d+2}$. The only effect of the $dt^2$ term in the metric is to lead to a shift of the mass term from $M_0^2$ to $M_0^2 + M^2$.} of mass $M_0$, one can see that the
dimensions of the dual operators are
\cite{Son:2008ye,Balasubramanian:2008dm} \beq \label{dimm} \Delta_M
= \frac{d+1}{2} + \sqrt{M_0^2 + M^2 + \frac{(d+1)^2}{4}}. \eeq In
the large $M$ limit this just becomes $\Delta \sim M$. So instead of
having a single operator one has a whole tower $O_M$. The dimensions
of the higher $O_M$ are not simple multiples of $O_0$. The fact that
all the $\Delta_M$ as given by (\ref{dimm}) are determined in
terms of $M$ and a single number $M_0$ is an artifact of using the
free wave equations. Higher order terms in the Lagrangian (in particular a
coupling of $\phi_{d+2}$ to the massive vector that is typically
used as the matter to support the Schr\"odinger geometry as a
solution to Einstein's equations) will make the dimensions of the
various $O_M$ completely independent. As a consequence, once one
constructs the gravity dual for a NRdCFT, one should no longer
expect the mode decomposition in the bulk to represent a single dOPE
but instead it has to yield the dOPE for every $O_M$; again one
would not expect the different $O_M$ that are dual to a given scalar
to have dOPEs that are related in an obvious way. What will be new
in the case of a dCFT is that we will see this independence of the
dOPEs of different $O_M$ in the same tower of operators dual to the
scalar field $\phi_{d+2}$ already for the simple example of a free
scalar field.

\subsection{Gravitational description of the non-relativistic dOPE}

A $d$-dimensional NRdCFT preserves a subgroup of the
non-relativistic conformal algebra that is equivalent to the
non-relativistic conformal algebra in $d-1$ dimensions. As in the
relativistic case, the difference between a $d$ dimensional
NRdCFT and a $d-1$ dimensional NRCFT is that the action of some of
the generators (in particular the dilatation operator) in the NRdCFT
do not just involve the coordinates on the defect, but also
transverse to the defect. When trying to construct a gravitational
background that respects the symmetries of a $d$ dimensional NRdCFT in
the same spirit as we did in the relativistic case (reviewed in
section \ref{cftgrav}), naively one would simply take a $d+2$
dimensional spacetime and slice it in terms of Sch$_{d+1}$ slices.
Just as in eq. (\ref{warpedads}) the Sch$_{d+1}$ on each slice in
the full metric is then multiplied by a warpfactor $e^{A(r)}$ that
is undetermined by symmetry considerations. However, as we just reviewed in
section \ref{nrcftgrav}, there are actually two independent
structures that can appear in the metric consistent with the $d-1$
dimensional non-relativistic conformal symmetry: $-dt^2/z^4$ and
AdS$_{d+1}$ in light cone coordinates (with a lightlike direction
compactified). Consequently we have two independent warpfactors in
the most general form of a metric that respects the symmetries of a
$d$-dimensional NRdCFT \beq \label{warpedsch} ds^2 = e^{2 A(r)}
\left ( - e^{2 B(r)} \frac{dt^2}{z^4} + \frac{-2 dv dt + d\vec{x}^2 +
dz^2 }{z^2} \right ) + dr^2. \eeq As in the relativistic case, we
know that far away from the defect the dCFT should recover the full
symmetry group of a $d$ dimensional NRCFT. That is, for large $r$ we
want the warped Sch$_{d+1}$ metric in (\ref{warpedsch}) to turn into
Sch$_{d+2}$. In order to see what condition this imposes on the
warpfactors $A(r)$ and $B(r)$ it is instructive to write Sch$_{d+2}$
in the warped Sch$_{d+1}$ form. Indeed it is easy to see that for
\beq e^{A(r)} = e^{B(r)} = \cosh(r) \eeq the same change of
variables (\ref{cov}) as in the relativistic case takes the metric
in (\ref{warpedsch}) into (\ref{schmetric})\footnote{To see this
note that $\cosh^4(r)dt^2/z^4$ directly turns into
$dt^2/\tilde{z}^4$ under (\ref{cov}). The remainder of the metric
then is just AdS$_{d+1}$ written in AdS$_d$ slicing and so we know
that (\ref{cov}) takes this into the standard flat slicing form
(\ref{flatslicing}) of AdS$_{d+1}$.}. So for a general NRdCFT,
asymptotically we have to have $A\sim B \sim r$.

As we saw in section \ref{nrdope}, the non-relativistic dOPE has
exactly the same form as the relativistic dOPE. Given this fact it
seems somewhat puzzling that symmetries allow the metric to contain
a second free function $B(r)$. If $A(r)$ encodes the dOPE, what
information does $B(r)$ carry? In order to shed light on this
question, we will once more look at the wave equation for a free
scalar of mass $M_0$. We make a mode-decomposition ansatz for the
field $\phi_{d+2}$ as in (\ref{ansatz}), \beq
\phi_{d+2}(t,\vec{x},v,z,r) = \sum_n \psi_n(r)
\phi_{d+1,n}(t,\vec{x},v,z). \eeq Plugging this into the mode
equation we again get \beq \label{laplonslicetwo} \nabla_{d+1}^2
\phi_{d+1,n}(t,\vec{x},v,z) = m_n^2 \phi_{d+1,n} (t,\vec{x},v,z),
\eeq where this time $\nabla_{d+1}^2$ is the Laplacian on
AdS$_{d+1}$. 
As in the non-defect case we
take the ansatz that $\phi_{d+1}(t,\vec{x},v,z) = e^{i M v}
\phi_{d}(t,\vec{x},z)$ to encode the particle number $M$ of the dual
operator.
The $m_n^2$ can again be obtained as eigenfunctions of a
radial equation which this time reads\footnote{Note that the only
non-trivial components of the inverse metric this time are
$g^{vv}=e^{2B-2A}, g^{vt} = g^{tv}=-z^2 e^{-2A}, g^{xx}=g^{zz}=z^2
e^{-2A}, g^{rr}=1.$ As a consequence, in an analogy with the non-defect
case, the only contribution from the $dt^2$ term in the metric is a shift of
the mass parameter proportional to $M^2$, but this time this shift has a non-trivial
$r$-dependence unless $A=B$. What is important for us is that in the separation of
 variables ansatz this term only affects the radial wave equation and so the
 $\phi_{d+1,n}$ can again be taken as eigenfunctions of the AdS$_{d+1}$ Laplacian. } \beq \psi_n''(r) + (d+1) A'(r) \psi_n'(r) +
e^{- 2 A(r)} m_n^2 \psi_n(r) - (M_0^2+M^2 e^{2B(r)-2A(r)}) \psi_n(r)
=0. \eeq The remaining steps in the derivation of the dOPE from the
mode expansion are identical to the relativistic case reviewed in
section \ref{cftgrav}, since the asymptotic change of variables
(given by (\ref{cov})) in the two cases is identical. Note that for
the special case $B=A$ the particle number $M$ simply leads to a
shift of the mass squared from $M_0^2$ to $M_0^2+M^2$ as in the
non-defect case. The mode equation then is basically the same as in
the relativistic case; the one difference is that the $A' \psi_n'$
term has a coefficient $d+1$ instead of the $d$ in the relativistic
case (as the additional $v$ direction contributes one additional
power of $e^A$ to the square root of the determinant of the metric).
The non-relativistic $d$ dimensional dCFT knows its origin from a
relativistic $d+1$ dimensional dCFT.

More interesting however is the case when $A \neq B$ (of course,
asymptotically they have to approach one another, but they can in
general be different functions of $r$). In this case the $M$-dependent mass shift is multiplied by a free function of $r$. Hence
the eigenvalues $m_n^2$ and the corresponding eigenfunctions will be
highly non-trivial functions of $M$, because $M$ doesn't simply give an
additive shift in the eigenvalue. Thus, we get completely different dOPEs for different
values of $M$. As we emphasized above, this was to be expected. Any
field in the NRCFT or NRdCFT setting is not dual to a single
operator, but to a whole tower of operators with particle number
$M$. Two free functions are possible in the defect geometry, since we are not just encoding one
dOPE, but really an infinite tower of
different dOPEs for an infinite tower of ambient space operators
$O_M$.

\section{The non-relativistic Janus solution: a specific example of an NRdCFT}
\label{holo}

In this section, we present an example of the holographic NRdCFTs discussed in previous sections, which we call the non-relativistic Janus solution.  The solution is constructed by applying the Null Melvin Twist to the Janus solution of type IIB supergravity.

\subsection{The relativistic Janus solution}
\label{Janus}

In \cite{Bak:2003jk}, an explicit domain-wall solution to the type IIB equations of motion was constructed.
The solution includes a non-trivial dilaton, metric and five-form field strength.  The geometry is asymptotically $AdS$, and therefore admits a dual field theory description.  The spacetime of the dual field theory is divided into two regions by a defect, and the coupling constant takes a different value in either half of the spacetime.  This two-faced nature of the field theory led the authors to call the solution the Janus solution.  Janus is non-supersymmetric, and stability is therefore a concern.  However, strong evidence for the stability of the solution was presented in \cite{Freedman:2003ax}.  The dual field theory was further investigated in \cite{Clark:2004sb}, where it was argued that the field theory was a particular deformation of $\mathcal{N} = 4$ SYM, with no matter fields living on the defect.  In this paper, our interest in Janus is that it provides an explicit solution to IIB supergravity with a dCFT dual, which we can, in turn, use to generate an explicit solution to IIB supergravity with an NRdCFT dual.

In our conventions, the metric, dilaton and five-form for the Janus solution in the Einstein frame are
\bea
ds^2_{E} & = & L^2 e^{2 A(\mu) } \( \frac{ -d\tau^2 + dx^2 + dy^2 + dz^2}{z^2} + d\mu^2 \) + L^2 ds^2_{S^5}, \\
\phi & = & \phi(\mu), \\
F & = & \frac{L^4 e^{5A(\mu)} }{ z^4} d\tau \wedge dx \wedge dy \wedge dz \wedge d\mu +\\
 && \frac{L^4}{8} \cos \tone \sin^3 \tone \sin \tthree ~d\chi \wedge d\tone \wedge d\ttwo \wedge d\tthree\wedge d\tfour.
\eea
We denote the time coordinate $\tau$, because it differs from the time coordinate of our NRdCFT.  We have chosen to represent the five-sphere as a Hopf fibration, as this allows for simple implementation of the Null Melvin Twist in the next section.  The explicit five-sphere metric in these coordinates is
\bea
ds^2_{S^5} & = & ds^2_{\mathbb{P}^2} + \( d\chi + \mathcal{A} \)^2 \\
& = & d\chi^2 + \sin^2\tone ~ d\chi d\ttwo +  \sin^2\tone \cos\tthree~ d\chi d\tfour + d\tone^2 + \frac{\sin^2\tone}{4} d\ttwo^2 \\
&& + \frac{ \sin^2\tone \cos \tthree}{2} d\ttwo d\tfour +
\frac{\sin^2 \tone}{4} d\tthree^2 + \frac{\sin^2\tone}{4} d\tfour^2.
\eea

One can check that the above ansatz solves the full type IIB supergravity equations of motion, provided that the dilaton and warpfactor obey
\bea
\label{phi'} \phi'(\mu) & = & c e^{-3A}, \\
\label{A'} A'(\mu) & = & \sqrt{ e^{2A} - 1 +\frac{c^2}{24} e^{-6A}
}.
\eea
Here, $c$ is a constant that sets the strength of the jump in coupling across the defect in the dual field theory.  The geometry is free from curvature singularities only if $c < \frac{9}{4\sqrt{2}}$.

\subsection{ The non-relativistic Janus solution}
\label{nrjanus}

The Schr\"odinger geometry was successfully embedded into IIB supergravity in \cite{Herzog:2008wg, Maldacena:2008wh, Adams:2008wt}.  The papers \cite{Herzog:2008wg, Adams:2008wt} accomplished this embedding through the use of the Null Melvin Twist (NMT).  The NMT \cite{Gimon:2003xk} is a series of boosts, T-dualities and twists that takes one solution to the IIB equations of motion and produces another.  In particular, the NMT takes a geometry that is asymptotically $AdS$ and transforms it into a geometry that is asymptotically Schr\"odinger.  For this reason, it is a powerful tool to construct holographic duals to NRCFTs using preexisting supergravity solutions that are asymptotically $AdS$.

The details of the NMT are given in Appendix \ref{NMT}.  Implementing the NMT requires the choice of two translational isometries.  It is for this reason that we chose to write the spherical metric as a Hopf fibration in the previous subsection.  Such a coordinate choice naturally identifies a $U(1)$ isometry on the sphere, and we choose to perform our twist in this direction.  Note that the choice of the $U(1)$ fiber breaks the symmetry of the sphere down to $U(1) \times SU(3)$.  This is reflected in the Melvinized Janus solution below.

Applying the NMT to the Janus solution is straightforward but tedious.  The details of this calculation and the conventions we used are collected in Appendix \ref{melvindetails}.   The resulting solution, in
Einstein frame, is
\bea
\label{NRJanus}
ds^2_E& = & L^2 e^{2A} \( -\frac{e^{2A} e^{\phi} }{ z^4} dt^2 + \frac{ -2dt dv + dx^2 + dz^2}{z^2} + d\mu^2 \) + L^2 ds^2_{S^5} \\
\phi & = & \phi(\mu) \\
B & = & \frac{ L^2 e^{2A} e^{\phi} }{ z^2} \(d\chi + \mathcal{A}\) \wedge dt \\
F & = &  \frac{L^4 e^{5A(\mu)} }{ z^4} dt \wedge dv \wedge dx \wedge dz \wedge d\mu +\\
 && \frac{L^4}{8} \cos \tone \sin^3 \tone \sin \tthree ~d\chi \wedge d\tone \wedge d\ttwo \wedge d\tthree\wedge d\tfour.
\eea
Again, one can check that, as long as the derivatives of $\phi$ and $A$ are still given by
 \bea
 \phi'(\mu) & = & c e^{-3A}, \\
A'(\mu) & = & \sqrt{ e^{2A} - 1 +\frac{c^2}{24} e^{-6A} },
\eea
this is a solution to the IIB equations of motion.

We see that \ref{NRJanus} does indeed take the expected NRdCFT form, as described in \ref{warpedsch}.\footnote{To have precise agreement with \ref{warpedsch}, one must enact a coordinate change from $\mu$ to $r$.}  From the coefficient of $dt^2$, we see that $B(\mu) = A(\mu) + \frac{\phi(\mu)}{2}$.  Also, while the five-sphere has remained intact in the metric, the presence of the one-form $d\chi + \mathcal{A}$ breaks the spherical symmetry as described above.

We have thus constructed a particular example of the holographic NRdCFTs outlined in earlier sections of this paper.  The dual NRdCFT is expected to consist of two regions of the twisted $\mathcal{N} = 4$ described in \cite{Herzog:2008wg}.  The two regions are separated by a codimension one defect, across which the coupling constant of the theories jumps.  Unfortunately, no matter fields live on the defect, so the non-relativistic Janus solution is not a viable system to create $p$-wave superfluids as described in \cite{Nishida:2008gk}.  NRdCFTs with matter living on the defect could be constructed in a similar fashion by beginning with a gravity setup involving probe branes, but we do not study that case in this paper.

\section{Conclusions and future directions}\label{conclude}

In this paper, we have begun an investigation into the holographic description of non-relativistic defect conformal field theories.  We have found that the defect Schr\"odinger symmetry allows for the presence of an arbitrary function of the radial coordinate in the metric.  This function does not affect the structure of the dOPE, which, in terms of defect operators, is the same as in the relativistic case.  However, the arbitrary function does play a crucial role in the eigenvalue equation determining the dimensions of defect operators appearing in the dOPE for a ambient operator of fixed particle number.  This feature helps elucidate how, in non-relativistic holography, a single bulk field is dual to a tower of boundary operators with differing particle number.

We have also managed to construct a particular example of an NRdCFT, which we have called the non-relativistic Janus solution.  This solution was achieved by applying the Null Mevlin Twist to the relativistic Janus solution of type IIB supergravity.  As discussed above, the non-relativistic Janus solution corresponds to a very particular choice of the arbitrary function allowed by the symmetries.  The corresponding boundary theory has no matter localized to the defect.  In this way, our specific example differs from the setups considered by Nishida in order to create $p$-wave superfluids.

The particularity of our solution naturally leads to the question of whether other NRdCFT solutions to IIB supergravity can be found.  We will now present a general ansatz consistent with our symmetries and the accompanying equations of motion.  The construction of a solution to these equations, other than the non-relativistic Janus solution, is left to future research.

The most general metric consistent with the defect Schr\"odinger symmetry was written down earlier in \ref{warpedsch}.  For a $(2 + 1)$-dimensional dual theory, the metric will read
\beq
ds^2 = e^{2 A(r)}
\left ( - e^{2 B(r)} \frac{dt^2}{z^4} + \frac{-2 dv dt + dx^2 +
dz^2 }{z^2} \right ) + dr^2 + (d\chi + \mathcal{A})^2 + ds^2_{\mathbb{CP}_2}.
\eeq
The generators of the $1+1$-dimensional defect NRCFT correspond to the following isometries of the metric:
\bea
H&:& t = t' + a \\
P&:& x = x' + b \\
N&:& v = v' + c \\
M&:& x = -x' \\
D&:& t = \lambda^2 t', ~~ v = v', ~~ x = \lambda x', ~~ z = \lambda z' \\
K&:& x = x' + \beta t', ~~ v = v' + \beta x' + \frac{1}{2} \beta^2 t' \\
C&:& t = \frac{t'}{1 + \alpha t'},~~ v = v' - \frac{ \alpha \((x')^{2} + (z')^{2}\)}{2 \( 1 + \alpha t'\)}, ~~ x = \frac{x'}{1 + \alpha t'}, ~~ z = \frac{z'}{1+ \alpha z'}.
\eea
Note that there is also another discrete isometry of the metric,
\beq
T: t = -t',~~ v = -v'.
\eeq
This corresponds to a time-reversal symmetry in the boundary theory.  In the non-relativistic Janus solution, in order for this to be a symmetry transformation, we must simultaneously take $B_{(2)} = -B'_{(2)}$.  Thus, we will require our Ansatz to have this symmetry.  We also require that our solution preserves the $U(1) \times SO(4)$ isometry of the compact directions.

The $H, P, N, D$ symmetries require the complex scalar field present in IIB supergravity (see \cite{Schwarz:1983qr} for our supergravity conventions) to depend only on the radial direction,
\beq
\tau = \tau(r).
\eeq

Next, let us work out the possible terms in the two-form.  First of all, the $H, P$ and $N$ transformations require the coefficients of all components to depend only on $z$ and $r$. The $M$ transformation prevents any two-form with an index in the $x$ direction.  The $T$ transformation (remember that this includes $B \rightarrow -B)$ requires every component to have an index in either the $t$ or the $v$ direction (but not both).  Let us assume there is some component of $B$ with an index in the $v$ direction,
\beq
B_{\mu v} = f(r) d\mu \wedge dv.
\eeq
Under the transformation $K$, $dv$ will pick up a term that has a $dx$.  The only other differential that would have a $dx$ to cancel such a term is the $x$ direction, but we have already established that $B$ cannot have indices in the $x$ direction.  Thus, $B$ also cannot have components in the $v$ direction, and must therefore have one index in the $t$ direction.

The remaining possibilities are
\beq
B = f_1(r, z) dt \wedge dz + f_2(r, z) dt \wedge dr + f_3 (r, z) dt \wedge (d\chi + \mathcal{A} ),
\eeq
where the last term is the only possible term with indices in the compact directions that is consistent with the $SO(4)$ isometry of $\mathbb{CP}_2$.  Finally, the $D$ and $C$ transformations uniquely determine the $z$ dependence of the functions $f_i$.  Our final most general two-form thus takes the form
\beq
B = \frac{e^{2C(r)}}{z^3} dt \wedge dz + \frac{ e^{2D(r)}}{z^2} dt \wedge dr + \frac{ e^{2E(r)}}{z^2} dt \wedge ( d\chi + \mathcal{A} ).
\eeq

The remaining field in IIB supergravity is the self-dual five-form.  In order to be consistent with the symmetries and to be self-dual, it must take the same form as in the non-relativistic Janus solution,
\bea
F & = &   \frac{ e^{5A(r)} }{ z^4} dt \wedge dv \wedge dx \wedge dz \wedge dr +\\
 && \frac{1}{8} \cos \tone \sin^3 \tone \sin \tthree ~d\chi \wedge d\tone \wedge d\ttwo \wedge d\tthree\wedge d\tfour.
 \eea

 In order to simplify the search for a solution, one could demand that the axion vanish and that the two-form is real, as in the non-relativistic Janus solution.  In the resulting equations of motion, $C(r)$ and $D(r)$ decouple from the other functions.  The equations of motion from the Janus case carry over, with the addition of two more equations that must be satisfied by $B(r)$ and $E(r)$.  The equations (independent of C(r) and D(r), which could be set to zero) reduce to the following set:
 \bea
 \phi'(\mu) & = & c e^{-3A}, \\
A'(\mu) & = & \sqrt{ e^{2A} - 1 +\frac{c^2}{24} e^{-6A} } \\
0 & = & 8e^{2A} - 6 + 2 A' \phi'  - 2 E' \phi' -2 A' E'- 4 (E')^2 -2 E'' \\
0 & = & e^{2A + 2B} \left( 10 + 6 A' B' + 4(B')^2 + 4 B' \phi' + c^2 e^{-6A} + 2B'' \right) \nonumber \\
&& -e^{4E} \left( 4 + 8e^{2A} + 4(E')^2 + 4E'\phi' + c^2 e^{-6A} \right).
\eea
We have not been able to find a simple solution to these equations.

\textbf{Acknowledgments:} The authors thank Dam Son for helpful
discussions.  This work was supported in part by DOE grant
DE-FG02-96ER40956.  The work of EGT was also supported by
DE-FG02-00ER41132, and that of PS by Polish science grant NN202
105136 (2009-2011).

\begin{appendix}

\section{The Null Melvin Twist}
\label{NMT}

The Null Melvin Twist, first introduced in \cite{Gimon:2003xk}, is a solution-generating technique that takes a solution to supergravity as an input and produces a second solution to supergravity, generally containing some lightlike NS-NS flux.  The procedure involves six steps.  They are as follows:
1. Boost by $\gamma$ along a translationally invariant direction $y$.  \\
2. T-dualize along the $y$-direction.  \\
3. Choose a one-form (for us it will be $d\chi + \mathcal{A}$) and shift it by a constant amount $\alpha$ in the $y$ direction. \\
4. T-dualize back along the $y$ direction. \\
5. Boost back by $-\gamma$ in the $y$ direction.  \\
6. Take the limit $\gamma \rightarrow \infty$, $\alpha \rightarrow 0$ such that $\beta = \frac{1}{2} \alpha e^{\gamma}$ is held fixed.

In order to enact the T-dualities in steps two and four, we will use the Buscher rules \cite{Buscher:1987sk}, which apply to the fields of IIB supergravity in the string frame.  Under these rules, the fields in the NS-NS sector transform into new, primed fields as
\bea
g_{yy}' & = & \frac{1}{g_{yy}}, ~~~~ g'_{ay} = \frac{B_{ay}}{g_{yy}}, ~~~~ g'_{ab} = g_{ab} - \frac{ g_{ay}g_{yb} + B_{ay}B_{yb}}{g_{yy}}, \\
B'_{ay} & = &\frac{g_{ay}}{g_{yy}}, ~~~~ B'_{ab} = B_{ab} - \frac{ g_{ay} B_{yb} + B_{ay}g_{yb} }{ g_{yy}}, \\
\phi' & = & \phi - \frac{1}{2} \log {g_{yy}}.
\eea
Here, $a$ and $b$ are any direction other than $y$.

\section{The NMT of Janus}
\label{melvindetails}

In this appendix we explicitly implement the NMT on the Janus
solution. Our discussion closely follows \cite{Adams:2008wt}.

In order to use the Buscher rules as listed in the previous section, we must start with a solution in string frame.  Before carrying out the NMT, we must therefore multiply the Janus metric in section \ref{Janus} by $e^{\phi/2}$.

The five-form is unaffected by the NMT.  The argument is
exactly the same as in \cite{Adams:2008wt}. Thus, we will only be
concerned with the dilaton, the $B$ field, and the following
components of the string frame metric:
\beq
ds^2 = -\frac{L^2 e^{2 A}e^{\phi/2}}{z^2} d\tau^2 +\frac{L^2 e^{2 A}e^{\phi/2}}{z^2} dy^2 + L^2 e^{\phi/2}
\(d\chi + \mathcal{A}\)^2.
\eeq
All other components of the metric are unaffected by the NMT.  We will denote the original dilaton present in the Janus solution as $\phi_0$.

1. The first step of the NMT is to boost by $\gamma$ in the y
direction,
\beq
\tau \rightarrow \cosh \gamma ~ \tau - \sinh \gamma ~ y,
~~~ y \rightarrow \cosh \gamma ~ y - \sinh \gamma ~ \tau.
\eeq
This has no effect on our fields, as our metric is manifestly boost invariant
and the initial B field is zero.

2.  The second step is to T-dualize in the y direction.  Using the Buscher rules described in the previous section, the nontrivial transformations are:
\bea
g_{yy} & \rightarrow & \frac{1}{g_{yy} } = \frac{z^2}{L^2 e^{2A} e^{\phi/2} } \\
\phi & \rightarrow & \phi - \frac{1}{2} \log g_{yy} = \phi_0 - \frac{1}{2} \log \frac{L^2 e^{2 A}e^{\phi/2}}{z^2} .
\eea

After T-dualizing, our metric dilaton and B-field are \bea
ds^2&  = & -\frac{L^2 e^{2 A}e^{\phi/2}}{z^2} d\tau^2 + \frac{z^2}{L^2 e^{2A} e^{\phi/2} } dy^2 + L^2 e^{\phi/2} \(d\chi + \mathcal{A}\)^2 \\
\phi & = & \phi_0 - \frac{1}{2} \log \frac{L^2 e^{2 A}e^{\phi/2}}{z^2}  \\
B & = & 0. \eea

3.  The third step is to shift $ \chi \rightarrow \chi + \alpha y$,
where $\alpha$ is a constant.  Only the metric is affected by this
procedure.  Our fields are now
\bea
ds^2&  = & -\frac{L^2 e^{2 A}e^{\phi/2}}{z^2} d\tau^2 + \frac{z^2 + L^4 e^{2A} e^{\phi} \alpha^2}{L^2 e^{2A} e^{\phi/2} } dy^2 + 2 \alpha L^2 e^{\phi/2} dy\(d\chi + \mathcal{A}\) \nonumber \\ &&+ L^2 e^{\phi/2} \(d\chi + \mathcal{A}\)^2 \\
\phi & = & \phi_0 - \frac{1}{2} \log \frac{L^2 e^{2 A}e^{\phi/2}}{z^2}  \\
B & = & 0. \eea

4.  The next step is to T-dualize back along the y direction.  We
use the shorthand $\chi$ to represent indices along the direction $(d\chi + \mathcal{A})$.  Using the Buscher rules again, we find
\bea
g_{yy} & \rightarrow & \frac{ L^2 e^{2A} e^{\phi/2}} {z^2 + L^4 e^{2A} e^{\phi} \alpha^2 } \\
g_{\chi y} & \rightarrow & \frac{B_{\chi y}}{ g_{yy}} = 0 \\
g_{\chi \chi} & \rightarrow & g_{\chi \chi} - \frac{ g_{\chi y}^2} { g_{yy} } = \frac{ L^2 z^2 e^{\phi/2} } { z^2 + L^4 e^{2A} e^{\phi} \alpha^2} \\
\phi & \rightarrow & \phi_0 - \frac{1}{2} \log \frac{ z^2 + L^4 e^{2A} e^{\phi} \alpha^2 }{ z^2} \\
B_{\chi y} & \rightarrow & \frac{ g_{\chi y}}{g_{yy}} = \frac{
\alpha L^4 e^{2A} e^{\phi}}{z^2 + L^4 e^{2A} e^{\phi} \alpha^2}.
\eea
After this, our fields read
\bea
ds^2&  = & -\frac{L^2 e^{2 A}e^{\phi/2}}{z^2} d\tau^2 + \frac{ L^2 e^{2A} e^{\phi/2}} {z^2 + L^4 e^{2A} e^{\phi} \alpha^2 }dy^2  +  \frac{ L^2 z^2 e^{\phi/2} } { z^2 + L^4 e^{2A} e^{\phi} \alpha^2} \(d\chi + \mathcal{A}\)^2 \\
\phi & = & \phi_0 - \frac{1}{2} \log \frac{ z^2 + L^4 e^{2A} e^{\phi} \alpha^2 }{ z^2}  \\
B & = & \frac{\alpha L^4 e^{2A} e^{\phi}}{z^2 + L^4 e^{2A} e^{\phi} \alpha^2}
\(d\chi + \mathcal{A}\) \wedge dy.
\eea

5.  The fifth step is to boost back by $-\gamma$ in the y direction,
\beq
\tau \rightarrow \cosh \gamma ~ \tau + \sinh \gamma ~ y, ~~~ y
\rightarrow \cosh \gamma ~ y + \sinh \gamma ~ \tau.
\eeq
After this boost, our fields are
\bea
ds^2&  = & -\frac{ L^2 e^{2A} e^{\phi/2} }{z^2} \frac{ z^2 + L^4 e^{2A} e^{\phi} \alpha^2 \cosh^2 \gamma }{z^2 + L^4 e^{2A} e^{\phi} \alpha^2 } d\tau^2 - \frac{2 L^6 e^{4A} e^{3\phi/2} \alpha^2 \cosh \gamma \sinh \gamma}{z^2 \( z^2 + L^4 e^{2A} e^{\phi} \alpha^2\) } dy d\tau \nonumber \\
&&+ \frac{ L^2 e^{2A} e^{\phi/2} }{z^2} \frac{ z^2 - L^4 e^{2A} e^{\phi} \alpha^2 \sinh^2 \gamma }{z^2 + L^4 e^{2A} e^{\phi} \alpha^2 } dy^2 + \frac{ L^2 z^2 e^{\phi/2} } { z^2 + L^4 e^{2A} e^{\phi} \alpha^2} \(d\chi + \mathcal{A}\)^2 \\
\phi & = & \phi_0 - \frac{1}{2} \log \frac{ z^2 + L^4 e^{2A} e^{\phi} \alpha^2 }{ z^2}  \\
B & = &  \frac{\alpha L^4 e^{2A} e^{\phi}}{z^2 + L^4 e^{2A} e^{\phi} \alpha^2}
\(d\chi + \mathcal{A}\) \wedge \( \cosh \gamma~ dy + \sinh \gamma ~
d\tau \).
\eea

6.  The final step is to take the limit $\alpha \rightarrow 0,
\gamma \rightarrow \infty$ such that $\frac{e^{\gamma} \alpha}{2} =
\beta$ is held fixed.  Functionally, this means wherever we have a
$\sinh \gamma$ or $\cosh \gamma$ times $\alpha$ we can replace it by
$\beta$, and wherever we just have an $\alpha$ we can drop those
terms.

After taking the limit, our fields become
\bea
ds^2&  = & -\frac{ L^2 e^{2A} e^{\phi/2} \( z^2 + L^4 e^{2A} e^{\phi} \beta^2\) }{ z^4 } d\tau^2 - \frac{2 L^6 e^{4A} e^{3\phi/2} \beta^2}{ z^4 } dy d\tau \nonumber \\
&&+ \frac{ L^2 e^{2A} e^{\phi/2} \( z^2 - L^4 e^{2A} e^{\phi} \beta^2 \)}{ z^4 }dy^2  +  L^2 e^{\phi/2} \(d\chi + \mathcal{A}\)^2 \\
\phi & = & \phi_0 \\
B & = & \frac{ \beta L^4 e^{2A} e^{\phi} }{ z^2  }  \(d\chi +
\mathcal{A}\) \wedge \( dy + d\tau \).
\eea

This completes the NMT procedure.  However, to make the
Schr\"odinger symmetry manifest, we will change to lightcone
coordinates,
\beq
\tau = \frac{1}{\sqrt{2} } \( t+ v\), ~~~ y = \frac{1}{\sqrt{2} } \( t - v\).
\eeq
In these coordinates, our fields are
\bea
ds^2&  = & -\frac{  2 \beta^2 L^6 e^{4A} e^{3\phi/2}}{z^4} dt^2  -\frac{ 2 L^2 e^{2A} e^{\phi/2}}{ z^2 } dt dv+  L^2 e^{\phi/2} \(d\chi + \mathcal{A}\)^2 \\
\phi & = & \phi_0 \\
B & = & \frac{ \sqrt{2} \beta L^4 e^{2A} e^{\phi} }{ z^2  }  \(d\chi +
\mathcal{A}\) \wedge \( dt \).
\eea
Finally, to clean things up, we make the change of coordinates
\beq
t \rightarrow \frac{t}{\sqrt{2} \beta L^2} ~~~ v \rightarrow \sqrt{2} \beta L^2 v,
\eeq
leaving
\bea
ds^2&  = & L^2 e^{2A} e^{\phi/2}  \( -\frac{ e^{2A} e^{\phi} }{z^4} dt^2  +\frac{-2 dt dv}{z^2} \)+  L^2 e^{\phi/2} \(d\chi + \mathcal{A}\)^2 \\
\phi & = & \phi_0 \\
B & = & \frac{ L^2 e^{2A} e^{\phi} }{ z^2  }  \(d\chi +
\mathcal{A}\) \wedge \( dt \).
\eea
Reverting back to Einstein frame, and restoring the unaffected components of the metric, we recover the non-relativistic Janus solution written down in \ref{nrjanus}.

\end{appendix}

\bibliographystyle{JHEP}
\bibliography{NRdCFT}

\end{document}